\documentclass{WileyMSP-template}

\usepackage{bm,color,amsmath,txfonts}
\usepackage{graphicx}
\usepackage{siunitx}
\usepackage{subfigure}
\usepackage{verbatim}
\usepackage{dcolumn}
\usepackage{bm}
\usepackage{epsf}
\usepackage{xcolor}
\usepackage{hyperref}
\usepackage{hhline}
\usepackage{float}
\usepackage{enumerate}
\usepackage{bbm}
\usepackage{lipsum}
\usepackage{cite}

\begin{document}

	\pagestyle{fancy}
	\rhead{\includegraphics[width=2.5cm]{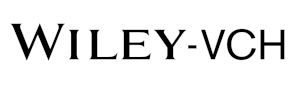}}

	\title{Microwave-optics entanglement via coupled opto- and magnomechanical microspheres}
	
	\maketitle

	
	\author{Hao-Tian Li},
	\author{Zhi-Yuan Fan},
	\author{Huai-Bing Zhu},
	\author{Simon Gr\"oblacher},
	\author{Jie Li}\thanks{jieli007@zju.edu.cn}

	\dedication{}

	\begin{affiliations}
		H.-T. Li, Z.-Y. Fan, H.-B. Zhu, J. Li\\
		Zhejiang Key Laboratory of Micro-Nano Quantum Chips and Quantum Control, School of Physics, and State Key Laboratory for Extreme Photonics and Instrumentation \\ 
		Zhejiang University\\
		Hangzhou 310027, China\\
		Email Address: jieli007@zju.edu.cn \\
		
		S. Gr\"oblacher\\
		Kavli Institute of Nanoscience, Department of Quantum Nanoscience\\
		Delft University of Technology\\
		2628CJ Delft, The Netherlands\\
	\end{affiliations}

	

	\begin{abstract}
		
		Microwave-optics entanglement plays a crucial role in building hybrid quantum networks with quantum nodes working in the microwave and optical frequency bands. However, there are limited efficient ways to produce such entanglement due to the large frequency mismatch between the two regimes.  Here, we present a new mechanism to prepare microwave-optics entanglement based on a hybrid system of two coupled opto- and magnomechanical microspheres, i.e., a YIG sphere and a silica sphere. The YIG sphere holds a magnon mode and a vibration mode induced by magnetostriction, while the silica sphere supports an optical whispering-gallery mode and a mechanical mode coupled via an optomechanical interaction. The two mechanical modes are close in frequency and directly coupled via physical contact of the two microspheres. We show that by simultaneously activating the magnomechanical (optomechanical) Stokes (anti-Stokes) scattering, stationary entanglement can be established between the magnon and optical modes via mechanics-mechanics coupling. This leads to stationary microwave-optics entanglement by further coupling the YIG sphere to a microwave cavity and utilizing the magnon-microwave state swapping. Our protocol is within reach of current technology and may become a promising new approach for preparing microwave-optics entanglement, which finds unique applications in hybrid quantum networks and quantum information processing with hybrid quantum systems.
		
	\end{abstract}
	

	\section{Introduction}
	\justifying
	Optical entanglement is a vital quantum resource, which finds broad and important applications in quantum information science and technology, such as quantum teleportation,\textsuperscript{\cite{bennett1993teleporting,bouwmeester1997experimental,furusawa1998unconditional}} quantum metrology,\textsuperscript{\cite{degen2017quantum,giovannetti2011advances}} quantum cryptology,\textsuperscript{\cite{beveratos2002single}} quantum logical operations,\textsuperscript{\cite{knill2001scheme}} fundamental tests of quantum mechanics,\textsuperscript{\cite{hensen2015loophole,giustina2015significant,shalm2015strong}} etc. To date,  many  ways have been proven to be efficient in producing entangled optical fields, e.g., by exploiting parametric down-conversion (PDC) in nonlinear crystals\textsuperscript{\cite{ou1992realization,Kwiat}} and linear optical operations such as beamsplitters.\textsuperscript{\cite{zukowski,kim02}} They have been generated in various systems, such as periodically poled lithium niobate waveguides,\textsuperscript{\cite{tanzilli2001highly}} quantum dots,\textsuperscript{\cite{benson2000regulated,stevenson2006semiconductor}} optical fibers,\textsuperscript{\cite{sharping2001observation,li2005optical}} atomic vapors,\textsuperscript{\cite{marino2009tunable}} coherent free electrons,\textsuperscript{\cite{baranes2022free}} etc.  In the microwave domain, entangled microwave fields are typically produced using Josephson parametric amplifiers.\textsuperscript{\cite{eichler2011observation,bergeal2012two,flurin2012generating}} They can also be generated by injecting squeezed vacuum through a microwave beamsplitter,\textsuperscript{\cite{menzel2012path}} reservoir engineering of an atomic beam,\textsuperscript{\cite{pielawa2007generation}} exploiting dynamical Casimir effect\textsuperscript{\cite{lahteenmaki2013dynamical}} and magnetostrictive interaction,\textsuperscript{\cite{yu2020magnetostrictively}} to name a few.
	
	Nevertheless, there are much fewer efficient ways to generate microwave-optics entanglement because of their large frequency mismatch. This, however, can be circumvented by coupling the optical and microwave fields to a common mechanical oscillator, utilizing the property of the mechanical oscillator that can couple to almost all electromagnetic fields. The optoelectromechanical system\textsuperscript{\cite{bagci2014optical,andrews2014bidirectional,forsch2020microwave,zhong2020proposal,barzanjeh2011entangling,zhong24,Painter20,Jiang20,Rob24}} then becomes a promising system to produce microwave-optics entanglement. This can be realized by simultaneously activating the electromechanical (optomechanical) PDC (beamsplitter) interaction, or alternatively the optomechanical (electromechanical) PDC (beamsplitter) interaction and using the mechanical oscillator as an intermediary to distribute the quantum correlation that is created in the PDC process. The microwave-optics entanglement can also be generated in a directly coupled electro-optics system via activating the PDC interaction,\textsuperscript{\cite{tsang2010cavity,rueda2019electro}} which has recently been demonstrated.\textsuperscript{\cite{sahu2023entangling}} Other approaches have also been proposed, e.g., using a cavity optomagnomechanical system.\textsuperscript{\cite{fan2023microwave}} 
	

	Here, we propose a novel approach for preparing stationary microwave-optics entanglement using a hybrid system combining opto- and magnomechanics. Specifically, a yttrium-iron-garnet (YIG) sphere and a silica sphere in physical contact are placed inside a microwave cavity. The YIG sphere holds a magnon mode and a mechanical vibration mode that are coupled via the magnetostrictive force.\textsuperscript{\cite{Zuo24}} The silica sphere supports an optical whispering-gallery mode (WGM) and a mechanical vibration mode that are coupled via the optomechanical interaction.\textsuperscript{\cite{aspelmeyer2014cavity}} The two mechanical modes are coupled via direct physical contact of the two spheres.\textsuperscript{\cite{shen22}} The magnon mode further couples to a microwave cavity mode via the magnetic dipole interaction. We use a red-detuned laser to drive the optical cavity, which activates the optomechanical anti-Stokes scattering and thus cools the mechanical mode of the silica sphere. Due to the mechanics-mechanics coupling, both mechanical modes can be effectively cooled to their quantum ground state at a low bath temperature. We further adopt a blue-detuned microwave field to drive the magnon mode, which activates the magnomechanical Stokes scattering and thereby the PDC interaction. Consequently, the magnomechanical entanglement is created, which is eventually distributed to the microwave and optical cavity modes through the effective microwave-magnon, mechanics-mechanics, and optomechanical beamsplitter interactions, giving rise to the microwave-optics entanglement. We identify that a {\it strong} mechanics-mechanics coupling is required to efficiently distribute the magnomechanical entanglement to the microwave and optical cavity modes.
	
	The paper is organized as follows. In Sec. \uppercase\expandafter{\romannumeral2}, we introduce the system under study consisting of two coupled opto- and magnomechanical microspheres and a microwave cavity. We provide the Hamiltonian and Langevin equations of the system, and show how to linearize the dynamics around the steady-state averages and obtain the covariance matrix in terms of quantum fluctuations, based on which entanglement is calculated. In Sec. \uppercase\expandafter{\romannumeral3}, we analyze the mechanism to produce microwave-optics entanglement in the system, present the results, and discuss some key issues that are vital for getting the entanglement. Finally, we summarize the findings in Sec. \uppercase\expandafter{\romannumeral4}.

	\section{The System}
	\justifying
	We consider an experimentally feasible system, as depicted in Figure 1a, which consists of a magnomechanical YIG sphere and an optomechanical silica sphere\textsuperscript{\cite{shen22}} that are in physical contact and placed inside a microwave cavity.  The magnon mode, e.g., the Kittel mode,\textsuperscript{\cite{kittel1948theory}} is embodied by the collective motion (spin wave) of a large number of spins in the YIG sphere. It is activated by placing the YIG sphere in a uniform bias magnetic field and applying a microwave drive field, e.g., via a loop antenna. Due to the large size of the YIG sphere, e.g., a 200-$\mu$m-diameter sphere used in Ref.,\textsuperscript{\cite{shen22}} a dispersive type interaction is dominant between the magnon mode (at GHz) and the magnetostriction induced mechanical mode (at $10^1$ MHz).\textsuperscript{\cite{Zuo24,zhang2016cavity,li2018magnon,potts2021dynamical,shen2022mechanical}}   The magnon mode further couples to a microwave cavity mode via the magnetic dipole interaction\textsuperscript{\cite{huebl2013high,tabuchi2014hybridizing,zhang2014strongly}} by placing the YIG sphere near the maximum magnetic field of the cavity mode (Note: the size of the cavity is much bigger than the YIG sphere and Figure 1a is only a diagrammatic sketch). 
	The silica sphere supports both an optical WGM and a mechanical mode coupled via the radiation pressure or photoelastic effect.\textsuperscript{\cite{aspelmeyer2014cavity}} Since the two microspheres are contacted closely, their localized mechanical modes establish a direct coupling. The size of the two spheres are carefully selected, such that the mechanical modes are close in frequency and they share a linear beamsplitter type coupling.\textsuperscript{\cite{shen22}} The Hamiltonian of such a hybrid system reads
	\begin{align}
		\begin{split}
			H / \hbar&=\sum_{\substack{j=a, m,c \\ b_1, b_2}} \omega_j j^{\dagger} j+g_{m a}\left(m^{\dagger} a+a^{\dagger} m\right)+g_{m b_1} m^{\dagger} m\left(b_1 + b_1^{\dagger}\right)\\
			&-g_{c b_2} c^{\dagger} c\left(b_2 + b_2^{\dagger}\right)+g_{b_1 b_2}\left(b_1^{\dagger} b_2+b_2^{\dagger} b_1\right)+H_{\rm dri} / \hbar,
		\end{split}
	\end{align}
	where $j\,{=}\,a, m, c, b_1, b_2$ $(j^\dagger)$ are the annihilation (creation) operators of the microwave cavity mode, the magnon mode, the optical cavity mode, and the two mechanical modes, respectively, satisfying the canonical commutation relation $[j, j^\dagger ] {=} 1$. Here $\omega_j$ are their corresponding resonant frequencies, $g_{ma}$ is the cavity-magnon coupling strength, $g_{b_1 b_2}$ is the mechanics-mechanics coupling strength between the two microspheres, and $g_{m b_1}$ $(g_{c b_2})$ is the bare magnomechanical (optomechanical) coupling strength, which can be significantly enhanced by driving the magnon mode (optical cavity) with a strong microwave (laser) field. The driving Hamiltonian $H_{\rm dri} / \hbar=i \Omega\left(m^{\dagger} e^{-i \omega_{d_1} t}-{\rm H.c.}\right)+i E\left(c^{\dagger} e^{-i \omega_{d_2} t}-{\rm H.c.}\right)$, corresponding to a microwave (laser) field applied to drive the magnon (optical cavity) mode. The Rabi frequency $\Omega=\frac{\sqrt{5}}{4} \gamma \sqrt{N} H_d$\textsuperscript{\cite{li2018magnon}} describes the coupling strength between the magnon mode and the drive magnetic field with frequency $\omega_{d_1}$ and amplitude $H_d$, where $\gamma$ is the gyromagnetic ratio, and $N$ is the total number of spins in the YIG sphere, and $E=\sqrt{2\gamma_cP_{L}/(\hbar\omega_{d_2})}$ denotes the coupling strength between the WGM and the driving laser through a fiber, where $P_{L}\left(\omega_{d_2}\right)$ is the power (frequency) of the laser field, and $\gamma_{c}$ is the decay rate of the WGM due to the coupling with the fiber.
	
	The above Hamiltonian leads to the following quantum Langevin equations (QLEs) by including the dissipations and input noises, which, in the interaction picture with respect to $\hbar\omega_{d_1}(a^{\dagger}a+m^{\dagger}m)+\hbar\omega_{d_2}c^{\dagger}c$, are given by
	\begin{align}\label{qles2}
		\begin{aligned}
			\dot{m}& =-(i\Delta_{m}+\gamma_{m})m-ig_{m b_1}(b_{1} + b^{\dagger}_{1})m-ig_{ma}a+\Omega+\sqrt{2\gamma_{m}}m_{in},  \\
			\dot{c}& =-(i\Delta_c+\gamma_c)c+ig_{c b_2}(b_{2}+ b^{\dagger}_{2})c+E+\sqrt{2\gamma_{c}}c_{in},  \\
			\dot{a}& =-(i\Delta_{a}+\gamma_{a})a-ig_{ma}m+\sqrt{2\gamma_{a}}a_{in},  \\
			\dot{b}_1& =-(i\omega_{b_{1}}+\gamma_{b_1})b_{1}-ig_{m b_1}m^{\dagger}m-ig_{b_1 b_2}b_{2}+\sqrt{2\gamma_{b_1}}b_{1,in},  \\
			\dot{b}_2& =-(i\omega_{b_2}+\gamma_{b_2})b_{2}+ig_{c b_2}c^{\dagger}c-ig_{b_1 b_2}b_{1}+\sqrt{2\gamma_{b_2}}b_{2,in} ,
		\end{aligned}
	\end{align}
	where $\Delta_{a(m)} \,{=}\,\omega_{a(m)}-\omega_{d_1}$, $\Delta_{c}\,{=}\,\omega_{c}-\omega_{d_2}$, $\gamma_j$ ($j=a, m, c, b_1, b_2$) is the dissipation rate of the corresponding mode, and $j_{in}$ are the input noise operators, which are zero-mean and obey the following correlation functions: $\langle j_{in}(t)j_{in}^{\dagger}(t^{\prime})\rangle=\left[N_{j}(\omega_{j})+1\right]\delta(t-t^\prime)$, and $\langle j_{in}^\dagger(t)j_{in}(t^{\prime})\rangle=N_j(\omega_j)\delta(t-t^{\prime})$. The mean thermal excitation number of each mode $N_{j}(\omega_{j})=\left[\exp(\hbar\omega_{j}/k_{B}T)-1\right]^{-1}$,  with $T$ being the bath temperature.
	
	The generation of microwave-optics entanglement requires sufficiently strong opto- and magnomechanical coupling strength, responsible for cooling the two low-frequency mechanical modes and creating magnomechanical entanglement, respectively. To this end, we apply two strong drive fields to the magnon and optical cavity modes, respectively, which lead to large amplitudes $|\langle m\rangle|, |\langle c\rangle|\gg1$. This enables us to linearize the system dynamics around the large average values by neglecting small second-order fluctuation terms. Consequently, we obtain a set of linearized QLEs for the quantum fluctuations, which can be written using quadratures and in the matrix form of
	\begin{align}
		\dot{u}(t)=Au(t)+n(t),
	\end{align}
	where $u(t)=\big[\delta X_{b_1}(t),\delta Y_{b_1}(t),\delta X_{b_2}(t),\delta Y_{b_2}(t),\delta X_{m}(t),\delta Y_{m}(t),$ $\delta X_{c}(t),\delta Y_{c}(t),\delta X_{a}(t),\delta Y_{a}(t) \big]^{T}$ is the vector of the quadrature fluctuations, $n(t)=\Big[\! \sqrt{2\gamma_{b_1}}X_{b_1}^{in}(t),\sqrt{2\gamma_{b_1}}Y_{b_1}^{in}(t),\sqrt{2\gamma_{b_2}}X_{b_2}^{in}(t),$ $\sqrt{2\gamma_{b_2}}Y_{b_2}^{in}(t), \sqrt{2\gamma_m}X_m^{in}(t),\sqrt{2\gamma_m}Y_m^{in}(t),$\\ 
	$\sqrt{2\gamma_c}X_c^{in}(t),\sqrt{2\gamma_c}Y_c^{in}(t),$ $\sqrt{2\gamma_a}X_a^{in}(t), \sqrt{2\gamma_a}Y_a^{in}(t) \Big]^T$ is the vector of the input noises, and the quadratures are defined as $X_{j}=\frac{1}{\sqrt{2}}(j+j^{\dagger})$ and $Y_j=\frac{i}{\sqrt{2}}(j^\dagger-j)$, and $\delta X_j$ and $\delta Y_j$ are the corresponding fluctuations. Similarly, the associated input noise operators $X_j^{in}$ and $Y_j^{in}$ can be defined. The drift matrix $A$ is given by
	\begin{eqnarray}    
		A=\begin{pmatrix}-\gamma_{b_1}&\omega_{b_1}&0&g_{b_1 b_2}&0&0&0&0&0&0\\-\omega_{b_1}&-\gamma_{b_1}&-g_{b_1 b_2}&0&0&-\sqrt{2}G_{m}&0&0&0&0\\0&g_{b_1 b_2}&-\gamma_{b_2}&\omega_{b_2}&0&0&0&0&0&0\\-g_{b_1 b_2}&0&-\omega_{b_2}&-\gamma_{b_2}&0&0&0&-\sqrt{2}G_{c}&0&0\\\sqrt{2}G_m&0&0&0&-\gamma_{m}&\tilde{\Delta}_m&0&0&0&g_{ma}\\0&0&0&0&-\tilde{\Delta}_m&-\gamma_{m}&0&0&-g_{ma}&0\\0&0&\sqrt{2}G_{c}&0&0&0&-\gamma_{c}&\tilde\Delta_{c}&0&0\\0&0&0&0&0&0&-\tilde\Delta_{c}&-\gamma_{c}&0&0\\0&0&0&0&0&g_{ma}&0&0&-\gamma_{a}&\Delta_{a}\\0&0&0&0&-g_{ma}&0&0&0&-\Delta_{a}&-\gamma_{a}\end{pmatrix}.
	\end{eqnarray}
	We have defined the effective magno- and optomechanical coupling strength: $G_m=-i\sqrt{2}g_{m b_1}\langle m\rangle$ and $G_{c}=i\sqrt{2}g_{c b_2}\langle c\rangle$. The
	steady-state averages of the magnon and optical modes are
	\begin{align}
		\left<m\right>=\frac{\Omega}{(i\tilde{\Delta}_{m}+\gamma_{m})+\frac{g_{ma}^{2}}{i\Delta_{a}+\gamma_{a}}}, \,\,\,\,\,  \left<c\right>=\frac{E}{(i\tilde{\Delta}_{c}+\gamma_{c})},
	\end{align}
	and the effective detunings $\tilde{\Delta}_{m}=\Delta_{m}+2g_{m b_1} {\rm Re} \langle b_{1} \rangle$ and $\tilde{\Delta}_{c}=\Delta_{c}-2g_{c b_2} {\rm Re} \langle b_{2}\rangle$, which include the frequency shift due to the mechanical displacement jointly caused by the photo- and magnetoelastic interactions.  The steady-state averages of the mechanical modes are 
	\begin{align}
		\begin{split}
			\left<b_{1}\right>&=\frac{\left|\left<c\right>\right|^{2}g_{c b_2}g_{b_1 b_2}-\left|\left<m\right>\right|^{2}g_{m b_1}(i\gamma_{b_{2}}-\omega_{b_2})}{g_{b_1 b_2}^{2}-(i\gamma_{b_{1}}-\omega_{b_1})(i\gamma_{b_{2}}-\omega_{b_2})} , \\
			\left<b_{2}\right>&=\frac{\left|\left<c\right>\right|^{2}g_{c b_2}(i\gamma_{b_{1}}-\omega_{b_1})-\left|\left<m\right>\right|^{2}g_{m b_1}g_{b_1 b_2}}{g_{b_1 b_2}^{2}-(i\gamma_{b_{1}}-\omega_{b_1})(i\gamma_{b_{2}}-\omega_{b_2})} .
		\end{split}
	\end{align}
	We note that the above drift matrix $A$ is derived under the optimal conditions for the microwave-optics entanglement, i.e., $|\Delta_a|,\, |\tilde{\Delta}_m|,\, |\tilde{\Delta}_c| \,\,{\simeq}\,\, \omega_{b_1} \,\,{\simeq}\,\, \omega_{b_2} \,\,{\gg}\, \, \gamma_j$, $j=a,m,c$ (Figure 1c), which we discuss later in Sec. \uppercase\expandafter{\romannumeral3}. These lead to the following approximate expressions: $\langle m\rangle\simeq -i\Omega/(\tilde\Delta_{m}-g_{ma}^{2}/\Delta_{a})$, and $\langle c\rangle\simeq-iE/\tilde{\Delta}_c$, which are pure imaginary numbers, and therefore the effective couplings $G_m$ and $G_c$ are approximately real.
	
	Due to the linearized dynamics and the Gaussian nature of the input noises, the steady state of the system is a five-mode Gaussian state, which can be characterized by a 10 $\times$ 10 covariance matrix (CM) $V$, of which the entries are defined as $V_{ij}=\langle u_i(t)u_j(t)+u_j(t)u_i(t)\rangle/2$ $(i,j=1,2,...,10)$. The steady-state CM can be obtained by directly solving the Lyapunov equation\textsuperscript{\cite{vitali2007optomechanical}}
	\begin{align}
		AV+VA^T=-D,
	\end{align}
	where $D=\mathrm{diag}\big[\gamma_{b_1}(2N_{b_1}+1),\gamma_{b_1}(2N_{b_1}+1),\gamma_{b_2}(2N_{b_2}+1),$
	$\gamma_{b_2}(2N_{b_2}+1),\gamma_{m}(2N_{m}+1),\gamma_{m}(2N_{m}+1),\gamma_{c}(2N_{c}+1),\gamma_{c}(2N_{c}+1),\gamma_{a}(2N_{a}+1),\gamma_{a}(2N_{a}+1)\big]$
	is the diffusion matrix and defined by $D_{ij}\delta(t-t^{\prime})=\langle n_i(t)n_j(t^{\prime})+n_j(t^{\prime})n_i(t)\rangle/2$. We adopt the logarithmic negativity\textsuperscript{\cite{vidal2002computable,adesso2004extremal,plenio2005logarithmic}} to quantify the microwave-optics entanglement, which is defined as
	\begin{align}
		E_{ca}=\max[0,-\ln(2\eta^-)],
	\end{align}
	where $\eta^{-}\equiv2^{-1/2}\left[\Sigma-(\Sigma^{2}-4\det V_{4})^{1/2} \right]^{1/2}$, $V_4$ is the 4 $\times$ 4 CM of the microwave and optical cavity modes, which is in the form of $V_{4}=\left[V_c,V_{ca};V_{ca}^{\rm T},V_a \right]$, with $V_c$, $V_a$ and $V_{ca}$ being the 2 $\times$ 2 blocks of $V_4$, and $\Sigma\equiv\det V_{c}+\det V_{a}-2\det V_{ca}$. Similarly, we can calculate the entanglement of any other two modes of the system.

	\section{Stationary Microwave-optics Entanglement}
	\justifying
	The hybrid system have two low-frequency mechanical modes, which have a large amount of thermal excitations even at low bath temperatures. The prerequisite of creating any quantum state in the system is to eliminate those thermal excitations and essentially all the modes are in or close to their quantum ground state.\textsuperscript{\cite{li2018magnon,li19}} This means that a mechanical cooling process must be present, which is realized in our system by activating the optomechanical anti-Stokes scattering via driving the WGM with a red-detuned laser field with $\tilde{\Delta}_{c}\simeq\omega_{b_2}$ (Figure 1c). We note that for a hybrid system with multi components,\textsuperscript{\cite{fan2023microwave,Fan23PRA}} the anti-Stokes scattering must be sufficiently strong such that all coupled modes are significantly cooled. To generate entanglement, the PDC interaction is typically required (despite of other unconventional mechanisms), which is achieved by activating the magnomechanical Stokes scattering, implemented by driving the magnon mode with a blue-detuned microwave field with $\tilde{\Delta}_{m}\simeq-\omega_{b_1}$ (Figure 1c). Note that the above cooling and PDC processes work optimally in the resolved sideband limit, where $\omega_{b_1} \simeq \omega_{b_2} \gg \gamma_{m}, \gamma_{c} $.\textsuperscript{\cite{aspelmeyer2014cavity,Zuo24}} There is a trade-off between the strength of the Stokes and anti-Stokes scattering, such that thermal excitations are eliminated, sizeable entanglement is produced, while the system remains stable.  To entangle the microwave and optical cavity modes, the PDC generated magnomechanical entanglement must be efficiently distributed or transferred to the two cavity modes. An ideal interaction for realizing quantum state transfer among different modes is the beamsplitter (state-swap) interaction.\textsuperscript{\cite{yu20,li21}} Fortunately, in our system the (microwave) cavity-magnon coupling, the mechanics-mechanics coupling, as well as the {\it effective} optomechanical coupling are all such type interaction, which enable the distribution of the magnomechanical entanglement finally to the two cavity modes (cf., Figure 1b).   Figure 2a confirms our analyses on the optimal detunings  $\tilde{\Delta}_{m}\simeq -\omega_{b_1}$ and $\tilde{\Delta}_{c}\simeq\omega_{b_2}$ for getting stationary microwave-optics entanglement. We further note that a high-efficiency magnon-to-microwave state transfer requires the two modes to be nearly resonant, $\Delta_{a} \simeq \tilde{\Delta}_{m}$, and strongly coupled, $g_{ma}> \gamma_m,\gamma_a$.\textsuperscript{\cite{yu20}} However, their coupling should not be too strong, otherwise will cause the two cavity-magnon polariton modes, formed by strongly coupled magnons and microwave cavity photons,\textsuperscript{\cite{huebl2013high,tabuchi2014hybridizing,zhang2014strongly}} to be off-resonant with the magnomechanical Stokes sideband due to the normal-mode splitting, which suppresses the Stokes scattering and thus reduces the degree of the entanglement, as shown by comparing Figure 2a with Figure 2b.
	
	Figure 2c (2d) shows the steady-state microwave-optics (magnon-optics) entanglement $E_{ac}$ ($E_{mc}$) versus the microwave-magnon and mechanics-mechanics couplings $g_{ma}$ and $g_{b_1 b_2}$. Clearly, the two figures exhibit a complementary relation, implying that the microwave-optics entanglement $E_{ac}$ is partially transferred from the magnon-optics system through the microwave-magnon beamsplitter coupling, and the magnon-optics entanglement $E_{mc}$ is a result of the magnomechanical entanglement $E_{mb_1}$ distributed to the magnon-optics system through the mechanics-mechanics-optics ($b_1$-$b_2$-$c$) path, cf., Figure 1b.   A high-efficiency quantum state transfer also requires the two mechanical modes to be nearly resonant, $\omega_{b_1} \simeq \omega_{b_2}$, and strongly coupled, $g_{b_1 b_2} >\gamma_{b_1}^{\rm eff}, \gamma_{b_2}^{\rm eff}$, with $\gamma_{b_j}^{\rm eff} \gg \gamma_{b_j} $ ($j=1,2$) being the increased effective mechanical damping rates due to the optomechancial cooling interaction and the mechanical beamsplitter coupling. Under the parameters of Figure 2, the two mechanical modes get significantly cooled and the linewidths are increased to be in the order of $\sim$1 MHz, as seen in the optical cavity output spectrum $S_{\rm c}^{\mathrm{out}}(\omega)$ in Figure 3a (see Appendix for detailed calculation of the output spectrum).  Therefore, when the coupling $g_{b_1 b_2}$ is larger than about 1 MHz, the mechanical system enters the strong-coupling regime accompanied with the normal-mode splitting in the spectrum as shown in Figures 3b-3d.  Similar to the microwave-magnon coupling, the mechanical coupling should also not be too strong, because a larger $g_{b_1 b_2}$ yields a larger splitting $2g_{b_1 b_2}$ of the two hybridized mechanical modes in the spectrum, which may lead the mechanical sidebands to be off-resonant with the magnon and cavity modes, i.e., away from the optimal conditions $-{\Delta}_{a} =-\tilde{\Delta}_{m} \simeq \tilde{\Delta}_{c} \simeq \omega_{b_1}\simeq\omega_{b_2}$ (Figure 1c). This is confirmed by Figure 2c: the entanglement decreases when $g_{b_1 b_2}$ continues to increase after exceeding the optimal value of about 2.4 MHz.

	Figure 2 is plotted with experimentally feasible parameters:\textsuperscript{\cite{Zuo24,shen22}} $\omega_{a,m}/2\pi=10$ GHz, $\omega_{b_1}/2\pi=20.15$ MHz, $\omega_{b_2}/2\pi=20.11$ MHz, optical cavity resonance wavelength $\lambda_c=1550$~nm, $\gamma_{a,m,c}/2\pi=1$ MHz, $\gamma_{b_1,b_2}/2\pi=100$ Hz, $G_m/2\pi=0.7$ MHz, $G_c/2\pi=2.7$ MHz, and $T=10$ mK. Under the parameters of Figure 2c and the optimal couplings $g_{ma}/2\pi=1.5$ MHz and $g_{b_1 b_2}/2\pi=2.4$ MHz, the effective mean phonon numbers of the two mechanical modes are $\bar{n}_{b_1}^{\mathrm{eff}}\simeq0.11$ and $\bar{n}_{b_2}^{\mathrm{eff}}\simeq0.08$. Clearly, they are cooled to their quantum ground state.  Note that a relatively strong optomechanical coupling $G_c/2\pi=2.7$ MHz is used for cooling, due to the multi dissipation channels of the hybrid system.\textsuperscript{\cite{fan2023microwave,Fan23PRA}} This corresponds to a laser power $P_{L}\simeq30$~mW for $g_{c b_2}/2\pi=100$ Hz.\textsuperscript{\cite{aspelmeyer2014cavity}} The magnomechanical coupling $G_m/2\pi=0.7$ MHz corresponds to a drive power $P_{0}\simeq4$ mW (drive magnetic field $H_d \simeq 3.3 \times 10^{-5}$~T) for $g_{m b_1}/2\pi=0.1$~Hz.\textsuperscript{\cite{Zuo24}} To determine the power, we use the relation between the drive magnetic field $H_d$ and the power $P_0$ via $H_d=(1/R)\sqrt{(2P_0\mu_{0}/\pi c)}$,\textsuperscript{\cite{li2018magnon}} where $\mu_0$ is the vacuum magnetic permeability, $c$ is the speed of the electromagnetic wave propagating in vacuum, and $R$ is the radius of the YIG sphere, which we take $R=100$ $\mu$m.   {Figure 4 shows the microwave-optics entanglement as a function of two effective couplings $G_m$ and $G_c$. Clearly, for a given $G_c$, a stronger magnomechanical coupling $G_m$ yields a larger degree of entanglement and the maximum entanglement corresponds to the maximum coupling $G_m^{\rm max}$ allowed by the stability condition. There is however an optimal optomechanical coupling $G_c$: When $G_c$ is too small, the mechanical modes are not efficiently cooled to their ground state; when $G_c$ is too large, it activates not only the optomechanical cooling interaction but also the PDC interaction~\cite{li2018magnon,vitali2007optomechanical}, of which the latter is detrimental in the present scheme because the PDC interaction (for creating entanglement) is already provided by the magnomechanical system. Under the parameters of Figure 4, the optimal couplings are $G_m/2\pi=1.4$ MHz and $G_c/2\pi=3.2$ MHz, corresponding to the maximum entanglement $E_{ca}\simeq0.17$.}
	
	In Figure 5a, we plot the stationary microwave-optics entanglement versus the bath temperature $T$ and the optical cavity decay rate $\gamma_c$. We take the microwave cavity and magnon decay rates $\gamma_{a,m}/2\pi=1$ MHz, which are the typical values in the cavity magnonic experiments.\textsuperscript{\cite{tabuchi2014hybridizing,zhang2014strongly}}  The entanglement is robust against the thermal noise and cavity decay rate as the entanglement can still be present for the temperature (the decay rate) up to $\sim$200 mK ($\sim$10 MHz) under the above realistic parameters. The entanglement is also very robust with respect to the mechanical damping rates and exists for the damping rates $\gamma_{b_1,b_2}$ up to $\sim$$10^5$ Hz at $T=10$ mK, as shown in Figure 5b. Finally, we remark that all the results of the entanglement (Figures 2, 4 and 5) are in the steady state, which is guaranteed by the negative eigenvalues (real parts) of the drift matrix.

	\section{Conclusion}
	We have presented a practical and efficient scheme to prepare stationary microwave-optics entanglement employing two coupled opto- and magnomechanical microspheres placed inside a microwave cavity. The mechanical modes are cooled to their quantum ground state by activating the optomechanical anti-Stokes scattering and utilizing the direct mechanical coupling between the two spheres.  The entanglement is generated by the magnomechanical PDC interaction and eventually distributed to the microwave and optical cavity modes through the effective microwave-magnon, mechanics-mechanics, and optomechanical state-swap interaction. We have analyzed the optimal conditions and identified the strong mechanical coupling between the two spheres for getting the maximal entanglement. The scheme is robust against various dissipations of the system and can be realized using the current technology. {Compared with the approaches based on other systems~\cite{bagci2014optical,andrews2014bidirectional,forsch2020microwave,zhong2020proposal,barzanjeh2011entangling,zhong24,Painter20,Jiang20,Rob24,tsang2010cavity,rueda2019electro,sahu2023entangling,fan2023microwave}, the present protocol based on two large-size YIG and silica spheres is more experiment-friendly and cost-saving as it does not require a nano-fabrication platform and the YIG and silica spheres can be purchased commercially. Our protocol can also be applied to planar configurations as long as the magnomechanical interaction remains a dispersive type~\cite{Zuo24}.} The proposed microwave-optics entanglement finds many important applications in hybrid quantum networks and quantum information processing with hybrid quantum systems.

	\medskip
	\section*{\textbf{Acknowledgements}} \par 
	
	This work was supported by National Key Research and Development Program of China (Grant No. 2022YFA1405200) and National Natural Science Foundation of China (Grant No. 92265202).
	
	\section*{APPENDIX: CALCULATION FOR OPTICAL CAVITY OUTPUT SPECTRUM}\label{appA}

	\setcounter{figure}{0}
	\setcounter{equation}{0}
	\setcounter{table}{0}
	\renewcommand\theequation{A\arabic{equation}}
	\renewcommand\thefigure{A\arabic{figure}}
	\renewcommand\thetable{A\arabic{table}}

	Here we provide the details on the calculation of the optical cavity output spectrum $S^{\rm{out}}_{\rm c}(\omega)$, from which we can identify the strong mechanics-mechanics coupling induced normal-mode splitting at the mechanical sidebands. Taking the Fourier transform of the QLEs~\eqref{qles2}, we obtain
	\begin{align}
		\begin{aligned}
			-i\omega m& =-(i\Delta_{m}+\gamma_{m})m-ig_{m b_1}(b^{\dagger}_{1}+b_{1})m-ig_{ma}a+\Omega 
			\\
			&+\sqrt{2\gamma_{m}}m_{in} ,  \\
			-i\omega c& =-(i\Delta_c+\gamma_c)c+ig_{c b_2}(b^{\dagger}_{2}+b_{2})c+E+\sqrt{2\gamma_{c}}c_{in} ,  \\
			-i\omega a& =-(i\Delta_{a}+\gamma_{a})a-ig_{ma}m+\sqrt{2\gamma_{a}}a_{in} , \\
			-i\omega b_1& =-(i\omega_{b_{1}}+\gamma_{b_1})b_{1}-ig_{m b_1}m^{\dagger}m-ig_{b_1 b_2}b_{2}+\sqrt{2\gamma_{b_1}}b_{1,in} , \\
			-i\omega b_2& =-(i\omega_{b_2}+\gamma_{b_2})b_{2}+ig_{c b_2}c^{\dagger}c-ig_{b_1 b_2}b_{1}+\sqrt{2\gamma_{b_2}}b_{2,in} .
		\end{aligned}
	\end{align}
	The above set of equations lead to the following linearized QLEs for the quantum fluctuations of the system by neglecting small second-order fluctuation terms:
	\begin{align}\label{AAA}
		\begin{aligned}
			-i\omega \delta m& =-(i\tilde{\Delta}_m+\gamma_m)\delta m+G_{m b_1}(\delta b_1^{\dagger}+\delta b_1)-ig_{ma}\delta a \\
			&+\sqrt{2\gamma_m}m_{in} , \\
			-i\omega \delta c& =-(i\tilde{\Delta}_c+\gamma_c)\delta c+G_{c b_2}(\delta b_2^{\dagger}+\delta b_2)+\sqrt{2\gamma_c}c_{in} , \\
			-i\omega \delta a& =-(i\Delta_a+\gamma_a)\delta a-ig_{ma}\delta m+\sqrt{2\gamma_a}a_{in} , \\
			-i\omega \delta b_1& =-(i\omega_{b_1}+\gamma_{b_1})\delta b_1+G_{m b_1}(\delta m^{\dagger}-\delta m)-ig_{b_1 b_2}\delta b_2\\
			&+\sqrt{2\gamma_{b_1}}b_{1,in},  \\
			-i\omega \delta b_2& =-(i\omega_{b_2}+\gamma_{b_2})\delta b_2+G_{c b_2}(\delta c^{\dagger}-\delta c)-ig_{b_1 b_1}\delta b_1\\
			&+\sqrt{2\gamma_{b_2}}b_{2,in},  \\
		\end{aligned}
	\end{align}
	where we have redefined the effective magno- and optomechanical couplings: $G_{m b_1}=-ig_{m b_1}\langle m\rangle$ and $G_{c b_2}=ig_{c b_2}\langle c\rangle$. By solving Eq.~\eqref{AAA}, we can obtain the quantum fluctuation of the optical cavity field $\delta c(\omega)$, which takes the form of
	\begin{align}
		\begin{aligned}
			\delta c(\omega)=\sum_{\substack{j=b_1,b_2\\m,c,a}}\left[A_{j}(\omega)j_{in}(\omega)+B_{j}(\omega)j_{in}^{\dagger}(-\omega) \right] ,
		\end{aligned}
	\end{align}
	where $A_{j}(\omega)$ and $B_{j}(\omega)$ are the coefficients associated with different input noises. This allows us to obtain the cavity output field $\delta c_{\rm{out}}(\omega)$ by using the input-output relation $\delta c_{\rm{out}}(\omega)=\sqrt{2\gamma_c}\delta c(\omega)-c_{\rm{in}}(\omega)$, and define the cavity output spectrum $S^{\rm{out}}_{\rm c}(\omega)$, i.e.,
	\begin{align}
		\begin{aligned}
			S_{\rm c}^{\mathrm{out}}(\omega)=\left< \delta c_{\rm{out}}(\omega)^{\dagger}\delta c_{\rm{out}}(\omega) \right>.
		\end{aligned}
	\end{align}
	By further using the input noise correlations in the frequency domain, we plot the cavity output spectrum in Figure 3 for different values of the mechanical coupling $g_{b_1 b_2}$.
	
	\medskip
	
	%


	\begin{figure}[h]
		\centering
		\includegraphics[width=0.65\linewidth]{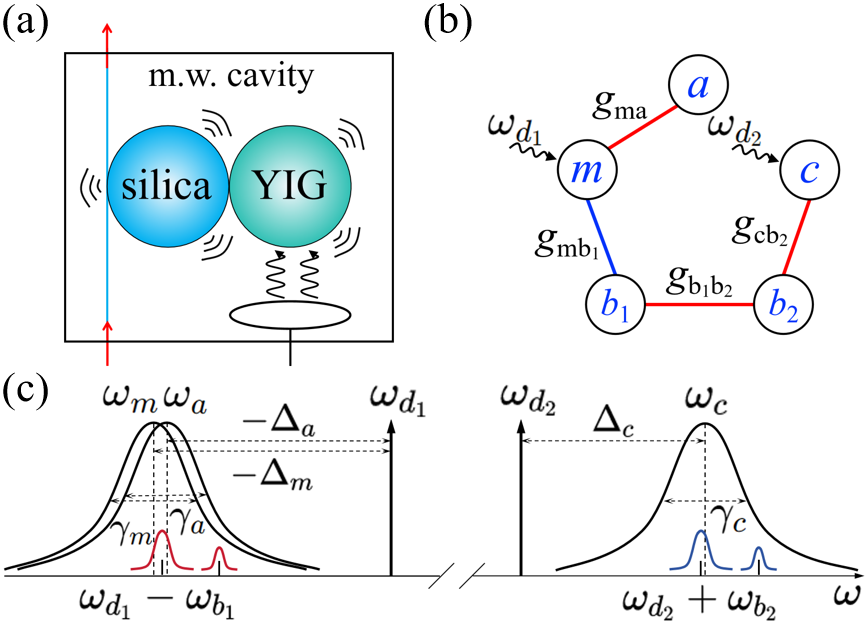}
		\caption*{Figure 1. a) Sketch of the hybrid system consisting of a YIG sphere and a silica sphere that are in physical contact and placed inside a microwave cavity. The YIG (silica) sphere supports a magnon (an optical) mode and a mechanical mode coupled via the magnetostrictive (optomechanical) interaction. The two localized mechanical modes of the two spheres are directly coupled due to their physical contact. The magnon mode further couples to the microwave cavity mode via the magnetic dipole interaction. b) Diagram of the couplings among different modes of the system. The blue (red) line denotes the effective PDC (beamsplitter) interaction. The magnon (optical) mode is strongly driven by a microwave (laser) field. c) Mode frequencies and linewidths used in the protocol. Two nearly resonant mechanical modes $\omega_{b_1}\simeq \omega_{b_2}$ are adopted to enhance the mechanical coupling. When the optical cavity is resonant with the  anti-Stokes sideband of the driving laser at $\omega_{d_2}+\omega_{b_2}$, and the magnon and microwave cavity modes are resonant with the Stokes sideband of the microwave drive field at $\omega_{d_1}-\omega_{b_1}$, stationary microwave-optics entanglement is established, which is maximized in the resolved sideband limit $\omega_{b_1} \simeq \omega_{b_2} \gg \gamma_m, \gamma_c$. Due to the close mechanical frequencies, there are also the Stokes sideband associated with the mechanical mode ($b_2$) at $\omega_{d_1}-\omega_{b_2}$ and the anti-Stokes sideband associated with the mechanical mode ($b_1$) at $\omega_{d_2}+\omega_{b_1}$.}
		\label{fig1}
	\end{figure}

	\begin{figure}[h]
		\includegraphics[width=0.65\linewidth]{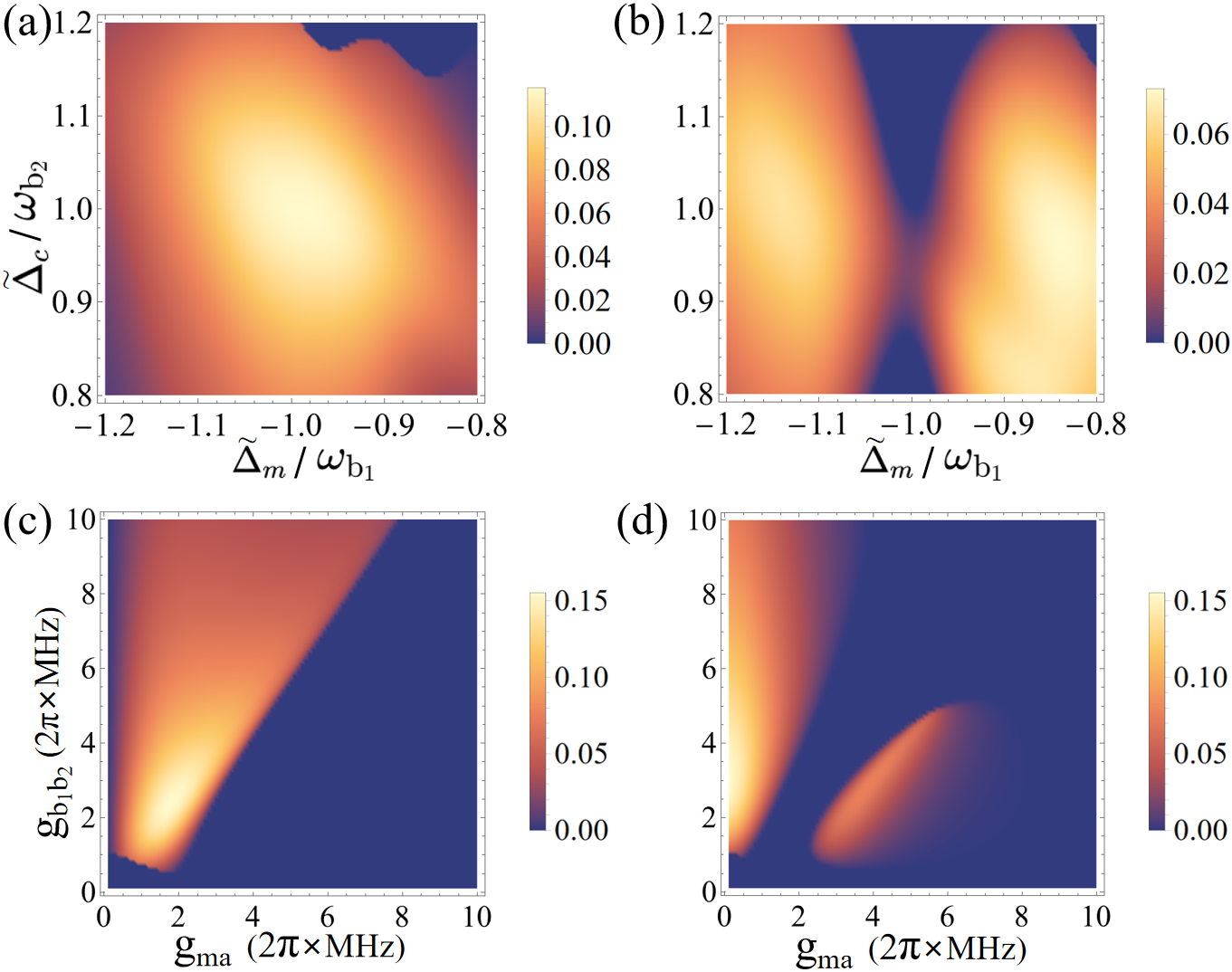}
		\caption*{Figure 2. Stationary microwave-optics entanglement $E_{ac}$ versus detunings $\tilde{\Delta}_{m} \,\, (=\Delta_{a})$ and $\tilde{\Delta}_{c}$ for a) $g_{ma}/2\pi=1$ MHz; b) $g_{ma}/2\pi=2.5$ MHz. In both plots, we take $g_{b_1 b_2}/2\pi$=1.5 MHz. c) Microwave-optics entanglement $E_{ac}$ and d) magnon-optics entanglement $E_{mc}$ versus coupling rates $g_{ma}$ and $g_{b_1 b_2}$ at the optimal detunings $\Delta_a=\tilde{\Delta}_{m}=-\omega_{b_1}$, and $\tilde{\Delta}_{c}=\omega_{b_2}$.  See text for the other parameters.}
		\label{fig2}
	\end{figure}

	\begin{figure}[h]
		\includegraphics[width=0.6\linewidth]{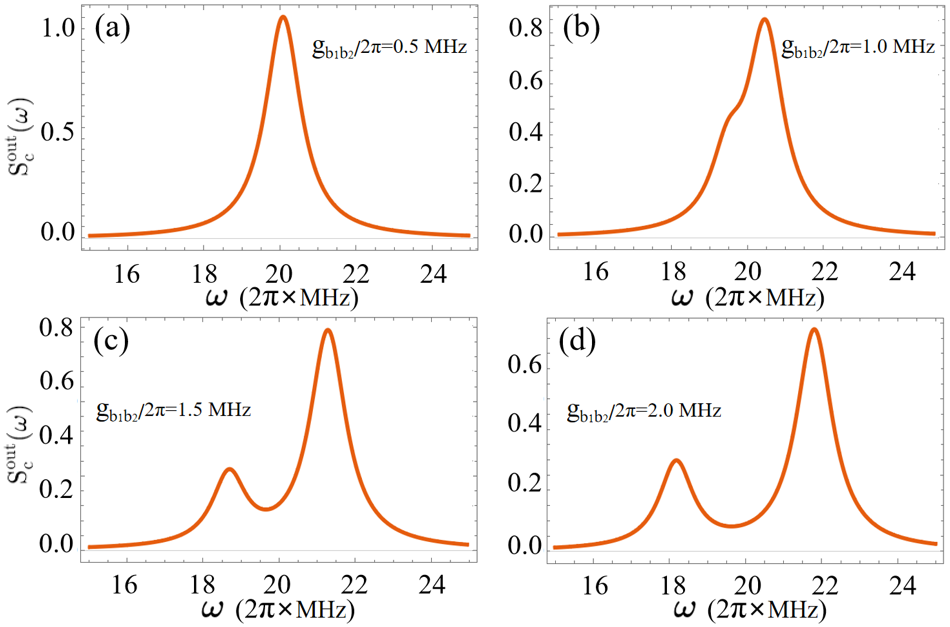}
		\caption*{Figure 3. Output spectrum of the optical cavity $S^{\rm{out}}_{\rm c}(\omega)$ with a) $g_{b_1 b_2}/2\pi=0.5$ MHz; b) $g_{b_1 b_2}/2\pi=1.0$ MHz; c) $g_{b_1 b_2}/2\pi=1.5$ MHz; d) $g_{b_1 b_2}/2\pi=2.0$ MHz. We take $g_{ma}/2\pi=1.5$ MHz, and optimal detunings $\Delta_a=\tilde{\Delta}_{m}=-\omega_{b_1}$ and $\tilde{\Delta}_{c}=\omega_{b_2}$. The other parameters are the same as in Figure 2.}
		\label{fig3}
	\end{figure}
	
	\begin{figure}[h]
		\includegraphics[width=0.33\linewidth]{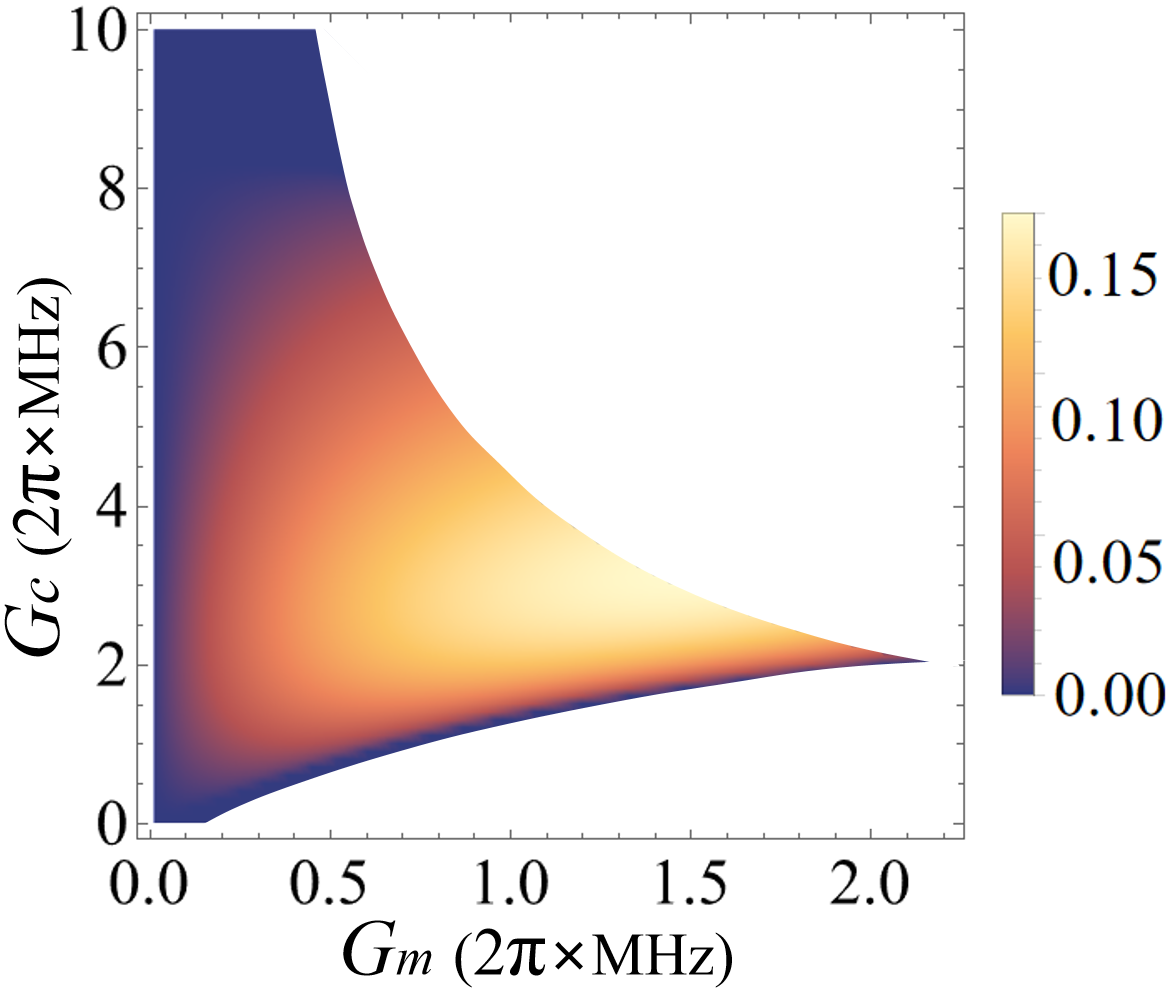}
		\caption*{Figure 4. Stationary microwave-optics entanglement $E_{ac}$ versus magnomechanical coupling strength $G_m$ and optomechanical coupling strength $G_c$. The blank area corresponds to the unstable regime. We take $\Delta_a=\tilde{\Delta}_{m}=-\omega_{b_1}$, $\tilde{\Delta}_{c}=\omega_{b_2}$, $g_{ma}/2\pi=1.5$ MHz, and $g_{b_1 b_2}/2\pi=2.4$ MHz. The other parameters are the same as in Figure 2.}
		\label{fig4}
	\end{figure}
	
	\begin{figure}[h]
		\includegraphics[width=0.67\linewidth]{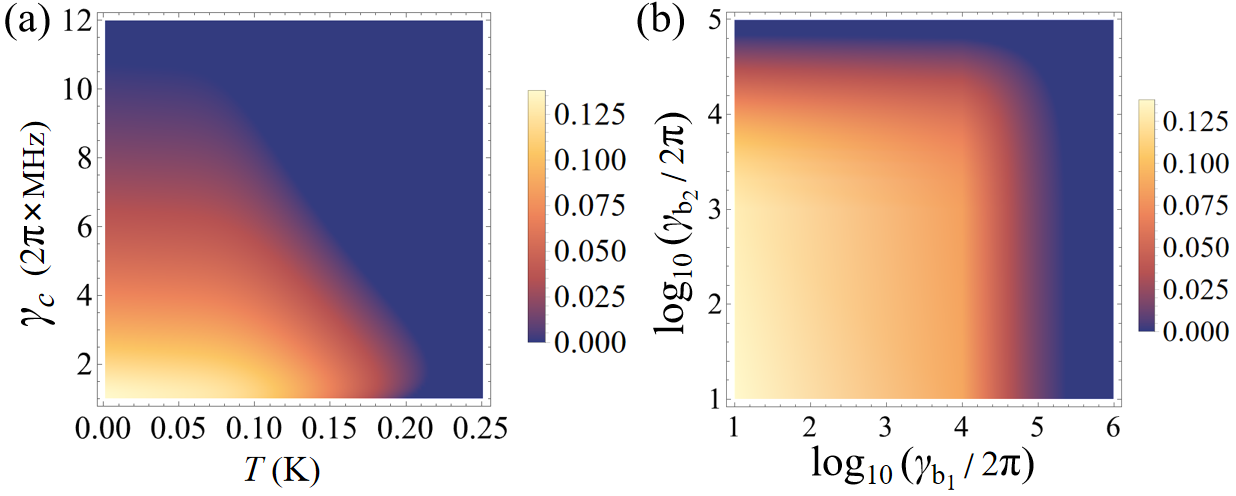}
		\caption*{Figure 5. Stationary microwave-optics entanglement $E_{ac}$ versus a) temperature $T$ and optical cavity decay rate $\gamma_c$; b) mechanical damping rates $\gamma_{b_1}$ and $\gamma_{b_2}$. We take $\Delta_a=\tilde{\Delta}_{m}=-\omega_{b_1}$, $\tilde{\Delta}_{c}=\omega_{b_2}$, $g_{ma}/2\pi=1.5$ MHz, and $g_{b_1 b_2}/2\pi=2.4$ MHz. The other parameters are the same as in Figure 2.}
		\label{fig5}
	\end{figure}

	\section*{\textbf{Table of Contents}} \par 
	Text:
	
	A new mechanism to generate microwave-optics entanglement is proposed using two coupled opto- and magnomechanical microspheres, i.e., a silica and a YIG sphere. Magnon-phonon entanglement is created in the YIG sphere by activating magnomechanical parametric down-conversion, which is distributed to the optical mode of the silica sphere through the mechanics-mechanics coupling and the optomechanical beamsplitter interaction, yielding stationary magnon-optics entanglement. This leads to stationary microwave-optics entanglement by further coupling the YIG sphere to a microwave cavity and exploiting magnon-microwave state swapping. The microwave-optics entanglement finds unique and important applications in hybrid quantum networks and quantum information processing with hybrid systems. \\
	

	
	\noindent Figure:
	\begin{figure}
		\includegraphics[width=0.3\linewidth]{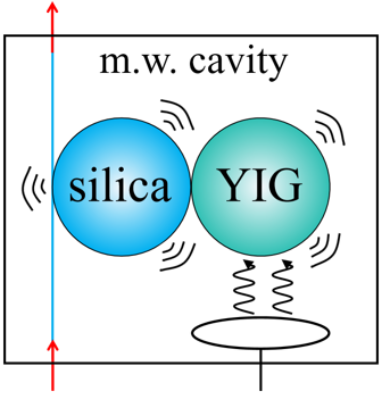}
		\caption*{ToC Figure Entry}
	\end{figure}

\end{document}